\documentclass[english,aps,showpacs,prb,twocolumn,floatfix,groupedaddress,a4paper]{revtex4-1}
\usepackage{amsmath}
\usepackage{graphicx}
\usepackage{amssymb}
\usepackage{color}
\usepackage{hyperref}
\usepackage{natbib}
\usepackage{dsfont}
\usepackage{units}
\usepackage{placeins}
\usepackage{bm}
\usepackage{dcolumn}

\makeatletter
\usepackage{babel}
\makeatother

\begin{document}
\title{Tunable $g$ factor and phonon-mediated hole spin relaxation in Ge/Si nanowire quantum dots}
\date{\today}
\author{Franziska Maier}
\affiliation{Department of Physics, University of Basel, Klingelbergstrasse 82, CH-4056 Basel, Switzerland}
\author{Christoph Kloeffel}
\affiliation{Department of Physics, University of Basel, Klingelbergstrasse 82, CH-4056 Basel, Switzerland}
\author{Daniel Loss}
\affiliation{Department of Physics, University of Basel, Klingelbergstrasse 82, CH-4056 Basel, Switzerland}
\begin{abstract}
We theoretically consider $g$ factor and spin lifetimes of holes in a longitudinal Ge/Si core/shell nanowire quantum dot that is exposed to external magnetic and electric fields. 
For the ground states, we find a large anisotropy of the $g$ factor which is highly tunable by applying electric fields. This tunability depends strongly on the direction of the electric field with respect to the magnetic field.
We calculate the single-phonon hole spin relaxation times $T_1$ for zero and small electric fields and propose an optimal setup in which very large $T_1$ of the order of tens of milliseconds can be reached. 
Increasing the relative shell thickness or the longitudinal confinement length further prolongs $T_1$. 
In the absence of electric fields, the dephasing vanishes and the decoherence time $T_2$ is determined by $T_2=2 T_1$. 
\end{abstract}
\maketitle
%
%
%%%%%%%%%%%%%%%%%%%%%%%%%%%%%%%%%%%%%%%%%%%%%%%%%%%%%%%%%%%%%%%
%
%
%
Semiconducting nanowires (NWs) allow to create nanoscale systems defined precisely regarding composition, geometry, and electronic properties and hence are subject to great experimental efforts. 
Furthermore, they offer new ways for implementing spin-based quantum computation. \cite{Loss1998}
Both III-V compounds and group-IV materials are considered and operated in the conduction band (CB, electrons) \cite{Doh2005,Fasth2007,Nilsson2009,NadjPerge2010,Schroer2011,NadjPerge2012,Petersson2012,vandenBerg2012} and in the valence band (VB, holes) \cite{Lauhon2002,Lu2005,Xiang2006,Xiang2006a,Hu2007,Roddaro2007,Roddaro2008,Hao2010,Yan2011,Nah2012,Hu2012,Zhang2012,Pribiag2013} regime. 
A particularly favored material is InAs, where single-electron quantum dots (QDs) \cite{Fasth2007} and electrically controlled spin rotations \cite{NadjPerge2010,Schroer2011,Petersson2012} have been implemented.
Recently, qubits have also been implemented in InSb NW QDs,\cite{NadjPerge2012,vandenBerg2012,Pribiag2013}  a system for which extremely large electron $g$ factors have been found. \cite{Nilsson2009,NadjPerge2012} 
However, the strong hyperfine interaction in InAs and InSb is considered the dominant source for the short coherence times observed. \cite{NadjPerge2010,vandenBerg2012}
The latter may therefore be substantially prolonged in group-IV NWs that can be grown nuclear-spin-free.
In this context, Ge and Si have emerged as promising materials for nanoscale systems such as lateral QDs, \cite{Simmons2011,Prance2012,Maune2012,Shi2012} self-assembled QDs, \cite{Wang2007,Katsaros2010,Katsaros2011} cylindrical core/shell NWs, \cite{Lauhon2002,Lu2005,Xiang2006,Xiang2006a,Hu2007,Roddaro2007,Roddaro2008,Hao2010,Yan2011,Nah2012,Hu2012} and  ultrathin, triangular NWs. \cite{Zhang2012}

For applications in spintronics and quantum information processing, it can be advantageous to consider holes instead of electrons. 
Due to the $p$-wave symmetry of the Bloch states, holes experience a strong spin-orbit interaction (SOI) on the atomic level leading to an effective spin $J=3/2$ behavior. 
Hence spin and momentum are coupled strongly which allows efficient control of the hole spin by electrical means. 
Furthermore, hole spin lifetimes are prolonged in the presence of confinement. \cite{Bulaev2005,Heiss2007,Trif2009,Fischer2008,Fischer2010,Brunner2009} 

In Ge/Si core/shell NWs, the large VB offset leads to an accumulation of holes in the core. \cite{Lu2005,Park2010}
They form a one dimensional (1D) hole gas with an unusually large, tunable Rashba-type SOI, referred to as direct Rashba SOI (DRSOI).  \cite{Kloeffel2011}
This DRSOI makes Ge/Si core/shell NWs attractive candidates for quantum information processing via electric-dipole induced spin resonance, \cite{Golovach2006} and we mention that signatures of a tunable Rashba SOI were already deduced from magnetotransport experiments. \cite{Hao2010}
Experiments on gate defined QDs in this system revealed an anisotropy and confinement dependence of the $g$ factor. \cite{Roddaro2007,Roddaro2008}
Recently, singlet-triplet relaxation times in the range of several hundred microseconds were measured. \cite{Hu2012} 

%
%
%%%%%%%%%%%%%%%%%%%%%%%%%%%%%%%%%%%%%%%%%%%%%%%%%%%%%%%%%%%%%%%
%
%
%
In this work, we consider holes forming qubits in the energetically lowest states of longitudinal QDs in Ge/Si core/shell NWs.
We find the effective $g$ factor $g_{\text{eff}}$ of this subsystem which turns out to be strongly anisotropic and tunable by choosing the direction and magnitude of applied electric fields. 
For small electric fields, we perturbatively derive an effective subspace Hamiltonian and the according hole spin phonon coupling and calculate the hole spin relaxation rate $T_1^{-1}$.
At small Zeeman splittings $\hbar \omega$ we observe a $\omega^{7/2}$ proportionality of $T_1^{-1}$ which contrasts the $\omega^{5}$ behavior found for electrons in QDs. \cite{Khaetskii2001,Golovach2004,Trif2008,Kroutvar2004,Amasha2008, Dreiser2008}
The magnitude of $T_1^{-1}$ depends strongly on the direction of the magnetic field with respect to the wire.
For zero electric field, aligning the magnetic field perpendicular to the wire results in very long $T_1$ of the order of tens of milliseconds.
Directing the magnetic field along the wire results in a much shorter $T_1$. 
For both configurations, the dephasing is zero, hence the decoherence time is given by $T_2=2 T_1$. 
Applying small electric fields can enhance the relaxation rate by several orders of magnitude.
This effect depends strongly on the direction of the electric field with respect to the magnetic field. 
Long $T_1$ in the presence of electric fields are obtained when electric and magnetic fields are perpendicular to each other and perpendicular to the wire.
Moreover, we find that $T_1$ can be prolonged further by increasing the relative shell thickness and the longitudinal QD confinement.
Thus, we predict an optimal field geometry  for spin qubits in Ge/Si NWs that can be tested experimentally.

\begin{figure}[t]
\centering
\includegraphics[width=\columnwidth]{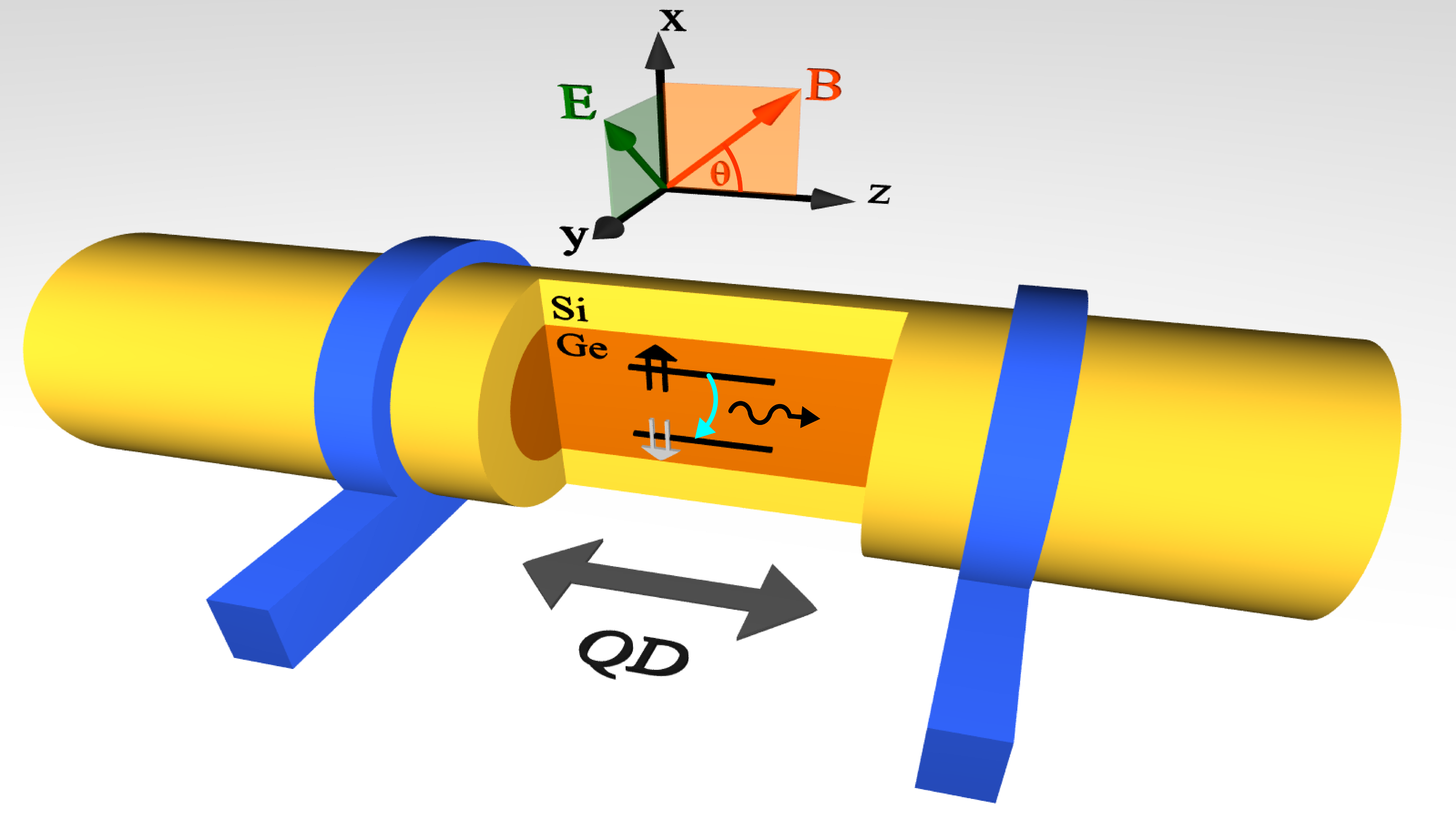}
\caption{Sketch of a Ge/Si core/shell NW aligned with the $z$ axis of the coordinate system. 
Electric gates (blue) induce confinement along the $z$ axis and define a QD. The electric field $\bm{E}$ lies perpendicular to the wire in the $xy$ plane and the magnetic field $\bm{B}$ lies in the $xz$ plane.
\label{fig:setupsystem}}
\end{figure}

%
%
%%%%%%%%%%%%%%%%%%%%%%%%%%%%%%%%%%%%%%%%%%%%%%%%%%%%%%%%%%%%%%%
%
%
%
Low-energetic hole states in a cylindrical Ge/Si core/shell NW are well described by an effective 1D Hamiltonian \cite{Kloeffel2011}
\begin{equation}
H_{w} = H_0 +H^{\prime} 
\end{equation}
that can be split into a leading order term $H_0$ and a perturbation $H^{\prime}$,
\begin{eqnarray}
H_0 &=& H_{\text{LK}_{\text{d}}} + H_\text{strain} + H_{B,Z},
\label{eq:LeadingOrder}\\
H^{\prime} &=& H_{\text{LK}_{\text{od}}}+H_{\text{R}}+H_{\text{DR}}+H_{B, \text{orb}}.
\label{eq:PerturbationOverview}
\end{eqnarray}
Using the notation introduced in Ref.~\onlinecite{Kloeffel2011} and defining the $z$ axis as the NW axis (see Fig.\ \ref{fig:setupsystem}), the diagonal terms of the Luttinger-Kohn (LK) Hamiltonian and the strain-induced energy splitting read
\begin{eqnarray}
H_{\text{LK}_{\text{d}}} + H_\text{strain} &=& A_{+}(k_z, \gamma) + A_{-}(k_z, \gamma) \tau_z . \label{eq:EffLKAndStrain}
\end{eqnarray}
Here, $\tau_i$ and $\sigma_i$ are the Pauli matrices for band index $(\{g,e\})$ and spin block $(\{+,-\})$ of the basis states $g_\pm(x,y)$ and $e_\pm(x,y)$ that provide the transverse motion. 
In Eq.\ (\ref{eq:EffLKAndStrain}), we defined $A_{\pm}(k_z, \gamma) \equiv \hbar^2 k_z^2(m_g^{-1} \pm m_e^{-1})/4\mbox{ $\pm$ } \Delta/2$, with $m_g\simeq m_0/(\gamma_1+2\gamma_s)$ and $m_e = m_0/(\gamma_1+\gamma_s)$ as the effective masses along $z$. 
Here, $\gamma_1$ and $\gamma_s$ are the Luttinger parameters in spherical approximation and $m_0$ denotes the bare electron mass. 
For Ge, $\gamma_1=13.35$ and $\gamma_s=5.11$. \cite{Lawaetz1970}
$\Delta \equiv \Delta_{\rm LK} + \Delta_{\rm strain}(\gamma)$ is the level splitting between the $g_\pm$ and $e_\pm$ states, $\gamma \equiv (R_s - R)/R$ is the relative shell thickness, and $R$ ($R_s$) is the core (shell) radius. 
The Zeeman coupling $H_{B,Z}$ with splitting $\hbar \omega_{B,Z}$ in the lowest-energy subspace ($g$ band) is determined by the magnetic field $\bm{B} = (B_x,0,B_z) \equiv  |\bm{B}|(\sin\theta, 0, \cos\theta)$ (Fig.~\ref{fig:setupsystem}), where we set $B_y = 0$ due to cylindrical symmetry.
the main contributions to $H^{\prime}$ are
\begin{eqnarray}
H_{\text{LK}_{\text{od}}} &=& C k_z \tau_y \sigma_x, \label{eq:EffLKod}\\
H_{\text{DR}} &=& e U (E_x \tau_x \sigma_z - E_y \tau_y ), 
\end{eqnarray}   
where $H_{{\text{LK}}_\text{od}}$ features the off-diagonal couplings with coupling constant $C=7.26 \hbar^2/(m_0 R)$ provided by the LK Hamiltonian as a consequence of the strong atomic level SOI.
$H_{\text{DR}}$ is the DRSOI that results from direct, dipolar coupling to an electric field $\bm{E} = (E_x, E_y, 0)$, where $U=0.15 R$. 
We note that $\hbar k_z = - i \hbar \partial_z$ in Eqs.\ (\ref{eq:EffLKAndStrain}) and (\ref{eq:EffLKod}) is the momentum operator along the wire. 
In the absence of longitudinal confinement the wave functions along $z$ are of type $e^{i k_z z}$ with $k_z$ as the wavenumber. 
$H_{\text{R}}$ is the conventional Rashba SOI, and, although fully taken into account in the present analysis, turns out to be negligible for the typical parameters and electric fields considered here. 
Finally, $H_{B, \text{orb}}$ denotes the orbital coupling to the magnetic field. 
Details on all elements of $H_0$ and $H^{\prime}$ are provided in Ref.~\onlinecite{Kloeffel2011} and in Eqs.\ (\ref{eq:HDot})-(\ref{eq:HBorb}) in the appendix.

%
%
%%%%%%%%%%%%%%%%%%%%%%%%%%%%%%%%%%%%%%%%%%%%%%%%%%%%%%%%%%%%%%%
%
%
%
We proceed with the derivation of an effective 1D Hamiltonian $H_{\text{h-ph}}$ for the coupling between low-energetic holes and acoustic phonons. There are three different types of acoustic phonon modes in cylindric NWs: torsional, dilatational, and flexural. \cite{Cleland2003}
We find four different modes $\lambda$ with dispersion relation $\omega_{\lambda}(q)$, where $q$ is the phonon wavenumber along the wire and the exact form of $\omega_{\lambda}$ depends strongly on the shell thickness.
For the torsional and dilatational mode ($\lambda = T,L$) $\omega_{\lambda}$ depends linearly on $q$, whereas for the two flexural modes ($\lambda = F_{\pm1}$) this dependence is quadratic.
The detailed derivation will be published elsewhere, in this work we directly apply the displacement field  $\bm{u}(\bm{r},t) = \sum_{\lambda,q}\left[\bm{u}_{\lambda}(q,\bm{r},t)b_{q,\lambda}(t)+\text{H.c.}\right]$ obtained for a finite shell following Refs.~\onlinecite{Cleland2003,Nishiguchi1994, Stroscio2001}. 
Here, $b_{q,\lambda}(t)=e^{-i \omega_{\lambda}(q) t}b_{q,\lambda}$ is the time-dependent phonon annihilation operator.
To derive $H_{\text{h-ph}}$, we insert the associated strain tensor components $\varepsilon_{ij}(\bm{r},t)$ in the Bir-Pikus Hamiltonian, \cite{BirPikus1974}
\begin{equation}
H_{\text{BP}} = b \left[\sum_i \varepsilon_{ii} J_i^2 +2 \left(\varepsilon_{xy}\{J_x, J_y\}+\text{c.p.}\right)\right],\label{eq:HBPspher}
\end{equation}
where we omitted the global shift in energy and used the spherical approximation.  
The $J_i$, $i=x,y,z$, are the effective spin-$3/2$ operators of the VB electrons and the anti-commutator is defined as $\{A,B\}=(A B+B A)/2$. 
For Ge, the deformation potential $b$ takes the value $b\simeq-2.5\mbox{ eV}$. 
\cite{BirPikus1974}
We finally obtain 
\begin{equation}
H_{\text{h-ph}} = \sum_{\lambda}H_{\lambda} = H_{\text{T}}+H_{\text{L}}+H_{F_{+1}}+H_{F_{-1}}
\end{equation}
by integrating out the transverse part of the matrix elements, i.e.\ by projecting the Hamiltonian onto the subspace spanned by $g_\pm$ and $e_\pm$. 
The components of $H_{\text{h-ph}}$ are given explicitly in Eqs.~(\ref{eq:HphonT})-(\ref{eq:Hphon-1}) in the appendix.

%
%
%%%%%%%%%%%%%%%%%%%%%%%%%%%%%%%%%%%%%%%%%%%%%%%%%%%%%%%%%%%%%%%
%
%
%
Longitudinal confinement is realized by electric gating (see Fig.~\ref{fig:setupsystem}), which is modeled by adding a harmonic confinement potential in the $z$ direction,
\begin{equation}
H_{qd} = H_{w}+V_{\text{c}}(z), 
\end{equation}
where $V_{\text{c}}(z) = \frac{1}{2} \alpha_c z^2$.
$H_{qd}$ describes the QD well if the longitudinal confinement length is much larger than $R$.
The basis states of $H_{qd}$ are products of type $g_\pm \psi^{g}_{m}$ and $e_\pm \psi^{e}_{m}$, where the $\psi^{g/e}_{m}(z)$ are eigenfunctions of the harmonic oscillator $\hbar^2 k_z^2 /(2 m_{g/e}) + V_{\text{c}}(z)$ and $m \in \{0,1,\ldots\}$ is the harmonic oscillator quantum number. 
The confinement energies $\hbar \omega_{g/e}$ relate to $\alpha_c$ via $\alpha_c= m_{g/e} \omega_{g/e}^2$ and the harmonic oscillator confinement lengths read $z_{g/e} = \sqrt{\hbar/(m_{g/e} \omega_{g/e})}$. 
 
%
%
%%%%%%%%%%%%%%%%%%%%%%%%%%%%%%%%%%%%%%%%%%%%%%%%%%%%%%%%%%%%%%%
%
%
%
From $H_{qd}$ we extract the effective $g$ factor $g_{\text{eff}}$ of the lowest-energy subsystem by performing an exact, numerical diagonalization which gives the Zeeman splitting $\Delta E_{Z,\text{num}}$ (defined as positive) and
\begin{equation}
g_{\text{eff}} = \frac{\Delta E_{Z,\text{num}}}{\mu_B |\bm{B}|},
\end{equation}
where $\mu_B$ denotes the Bohr magneton. 
\begin{figure}[t]
\centering
\includegraphics[width=\columnwidth]{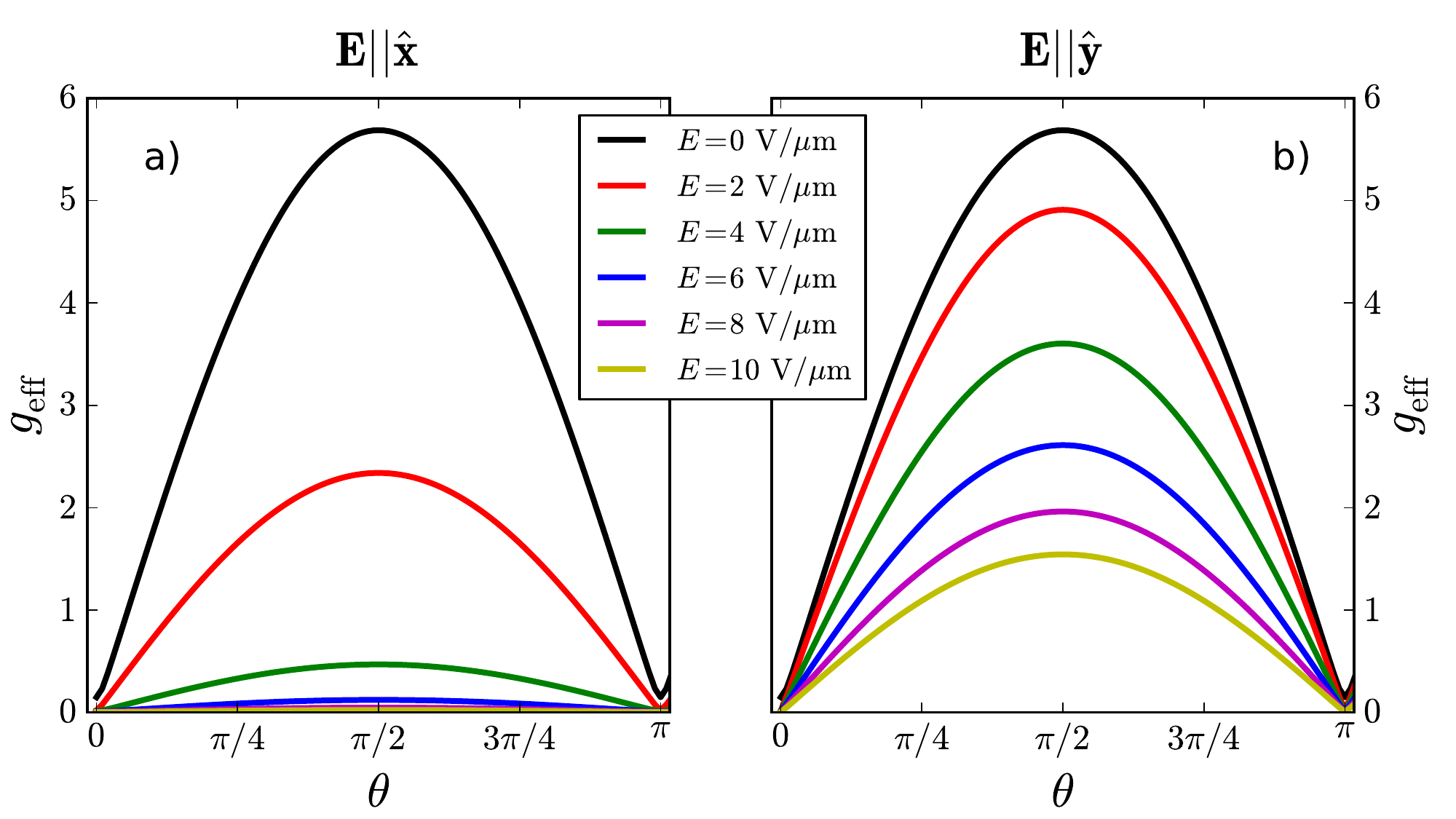}
\caption{Effective $g$ factor $g_{\text{eff}}$ as a function of the angle $\theta$ defined by $\bm{B}=|\bm{B}|(\sin\theta, 0, \cos\theta)$ for $\bm{E} \|\hat{\bm{x}}$ (a) and $\bm{E} \| \hat{\bm{y}}$ (b). 
We vary $|\bm{E}|$ from $0$ to $10\mbox{ {V}/$\mu${m}}$.
For $|\bm{E}|=0$, we find $g_{\text{eff}}(0)\approx 0.14$ and $g_{\text{eff}}(\pi/2)\approx 5.7$.  
It is clearly visible that $g_{\text{eff}}$ is affected much stronger by changes in $|\bm{E}|$ for $\bm{E} \|\hat{\bm{x}}$ than for $\bm{E} \| \hat{\bm{y}}$. 
Even though the curves in (a) seem to overlap for $|\bm{E}|\geq 6\mbox{ {V}/$\mu${m}}$, $g_{\text{eff}}(\pi/2)$ still decreases for growing fields and $g_{\text{eff}}$ remains anisotropic.
We choose $R=10 \mbox{ nm}$ and $R_s=13 \mbox{ nm}$ for the NW and a QD confinement length of $z_g\approx80\mbox{ nm}$. 
\label{fig:geffective}}
\end{figure}
In Fig.~\ref{fig:geffective}, we plot $g_{\text{eff}}$ as a function of the angle $\theta$, for both $\bm{E} \| \hat{\bm{x}}$ and $\bm{E} \| \hat{\bm{y}}$.
In both cases, $g_{\text{eff}}$ is highly anisotropic and tunable over a wide range of values by adjusting the magnitude of $\bm{E}$.
The tunability is caused by two mechanisms which occur in the system for large $|\bm{E}|$. 
The admixture of the $e_{\pm}$ states to the effective lowest-energy subsystem increases while the spin-orbit length $l_{\text{SOI}}$ decreases. 
For very small $l_{\text{SOI}}$ $(l_{\text{SOI}}\ll z_g)$, the hole spin flips many times while moving through the QD and the resulting $g_{\text{eff}}$ starts to average out.
The tunability is much stronger for $\bm{E} \| \hat{\bm{x}}$ than for $\bm{E} \| \hat{\bm{y}}$. 
Note that $g_{\text{eff}}$ is also tunable by varying $V_{\text{c}}(z)$. 
We find good agreement with the results given in Ref.~\onlinecite{Hu2012}, where $g_{\text{exp}}\approx 1.02$ was measured for $\bm{B}$ aligned with the NW with an accuracy of $\sim 30^{\circ}$. We note, however, that clearly different results for $g$ can be expected in QDs with very large occupation number, i.e.\ when the hole spin qubits are formed in an excited band.

%
%
%%%%%%%%%%%%%%%%%%%%%%%%%%%%%%%%%%%%%%%%%%%%%%%%%%%%%%%%%%%%%%%
%
%
%
In the following, we are interested in the dynamics of the lowest lying, Zeeman split states which we decouple perturbatively from the higher energy states. 
This is done by two consecutive Schrieffer-Wolff transformations (SWTs) to account for the two different energy scales $\Delta$ and $\hbar \omega_g$.
The general form of the SWT is $\tilde{H}=e^{-S}H e^S$, where to lowest order $S\approx S_1$. 
We first remove the coupling between the $g_{\pm}$ and $e_{\pm}$ states in the effective 1D picture using $S_{1}^g$.
The hole-phonon coupling then transforms according to $H_{\text{h-ph}}-[S_{1}^g,H_{\text{h-ph}}]$ and we refer to its projection on $g_{\pm}$ as $H^{g}_{\text{h-ph}}$.
In the second step, we add harmonic confinement as introduced above and decouple the two lowest, Zeeman split states $|0\rangle \equiv \{|\!\!\Uparrow\rangle,|\!\!\Downarrow\rangle\}$ by another SWT using $S_{1}^{|0\rangle}$.
A necessary condition for this approach is that the energy splittings obey $\Delta\gg\hbar \omega_g \gg \hbar \omega_{B,Z}$, and the magnitude of $\bm{E}$ is restricted by $2 C |\bm{E}| e U/(z_g \Delta) \ll \hbar \omega_g$.
The latter condition is fulfilled for $|\bm{E}|\ll1 \mbox{ {V}/$\mu${m}}$.
We obtain an effective Zeeman term $H_{Z,\text{eff}} = \mu_B \bm{B}_{\text{eff}} \cdot \bm{\sigma}$ with Zeeman splitting $\Delta E_{Z,\text{eff}} = 2 \mu_B |\bm{B}_{\text{eff}}|$, where $\bm{\sigma}$ is a vector of Pauli matrices.
The effective hole spin phonon coupling is obtained by taking
\begin{equation}
 H_{\text{s-ph}}=H^{g}_{\text{h-ph}}-[S_{1}^{|0\rangle},H^{g}_{\text{h-ph}}],
\end{equation}
where $H^{g}_{\text{h-ph}}$ is now written in the basis given by the confinement. 
Projecting $H_{\text{s-ph}}$ on $|0\rangle$ results in an effective coupling $H_{\text{s-ph},\text{eff}} = \mu_B \delta\bm{B} \cdot \bm{\sigma}$ with the fluctuating magnetic field $\delta\bm{B}(t) = \sum_{\lambda,q}\left[\bm{a}_{\lambda}(q) b_{q,\lambda}(t)+\text{H.c.}\right]$. 
The effective subspace Hamiltonian then reads
\begin{equation}
H_{\text{eff}} =H_{Z,\text{eff}}+H_{\text{s-ph},\text{eff}}  = \mu_B \left(\bm{B}_{\text{eff}} + \delta\bm{B}(t)\right) \cdot \bm{\sigma}. 
\label{eq:effcouplingHam}
\end{equation}
The spin relaxation rate in the Born-Markov approximation is given by the Bloch-Redfield approach \cite{Slichter1980,Golovach2004, Borhani2006} 
\begin{eqnarray}
\frac{1}{T_1}&=&n_i n_j \left[\delta_{ij}(\delta_{pq}-n_p n_q)J_{pq}^{+}(\omega)-(\delta_{ip}-n_i n_p)J_{pj}^{+}(\omega)\right.\nonumber\\
&&\left.-\delta_{ij}\varepsilon_{kpq}n_k I^{-}_{pq}(\omega)+\varepsilon_{ipq}n_p I^{-}_{qj}(\omega)\right],\label{eq:relaxrateStart}
\end{eqnarray}
where summation over repeated indices is assumed, $\bm{n} = \bm{B}_{\text{eff}}/|\bm{B}_{\text{eff}}|$ is the unit vector in direction of the effective magnetic field, and $\hbar\omega = \hbar \omega_{Z, \text{eff}} = \Delta E_{Z,\text{eff}}$ is the energy splitting of the considered states. 
Here, $J_{ij}^{+}(\omega)=\text{Re}[J_{ij}(\omega)+ J_{ij}(-\omega)]$ and $I_{ij}^{-}(\omega)=\text{Im}[J_{ij}(\omega)- J_{ij}(-\omega)]$, with $J_{ij}(\omega) = (\mu_B/\hbar)^{2}\int_0^{\infty}\mathrm{d}t e^{-i \omega t}\langle \delta B_i (0) \delta B_j (t)\rangle$ denoting the spectral function.
 
%
%
%%%%%%%%%%%%%%%%%%%%%%%%%%%%%%%%%%%%%%%%%%%%%%%%%%%%%%%%%%%%%%%
%
%
%
%
\begin{figure}[t]
\centering
\includegraphics[width=\columnwidth]{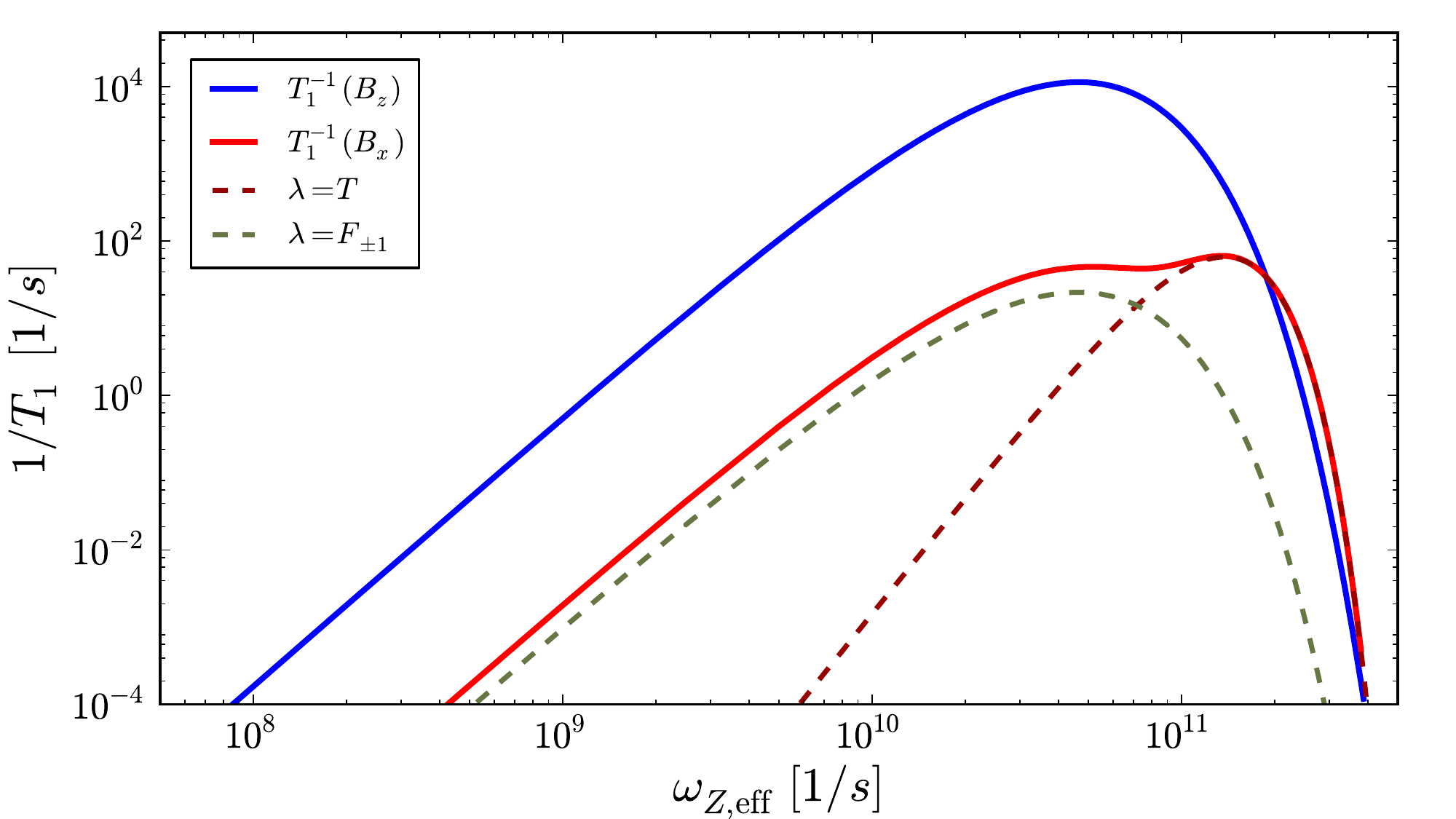} 
\caption{Relaxation rate $T_1^{-1}$ at $|\bm{E}|=0$ for $\bm{B}\|\hat{\bm{x}}$ (red, solid) and $\bm{B}\|\hat{\bm{z}}$ (blue, solid). 
For $\bm{B}\|\hat{\bm{x}}$, we plot the contributing phonon branches $F_{\pm1}$ and $T$ (dashed).
We find maximal values $T_{1,\text{max}}^{-1}(\bm{B}\|\hat{\bm{z}})\approx 11\mbox{ ms}^{-1}$ and $T_{1,\text{max}}^{-1}(\bm{B}\|\hat{\bm{x}})\approx 60\mbox{ s}^{-1}$. 
Note the non-monotonic behavior of $T_1^{-1}$ as function of $\omega_{Z,\text{eff}}$. 
The NW and QD parameters are chosen as in Fig.~\ref{fig:geffective}. 
\label{fig:relaxratezerofields}}
\end{figure}
\begin{figure}[t]
\centering
\includegraphics[width=\columnwidth]{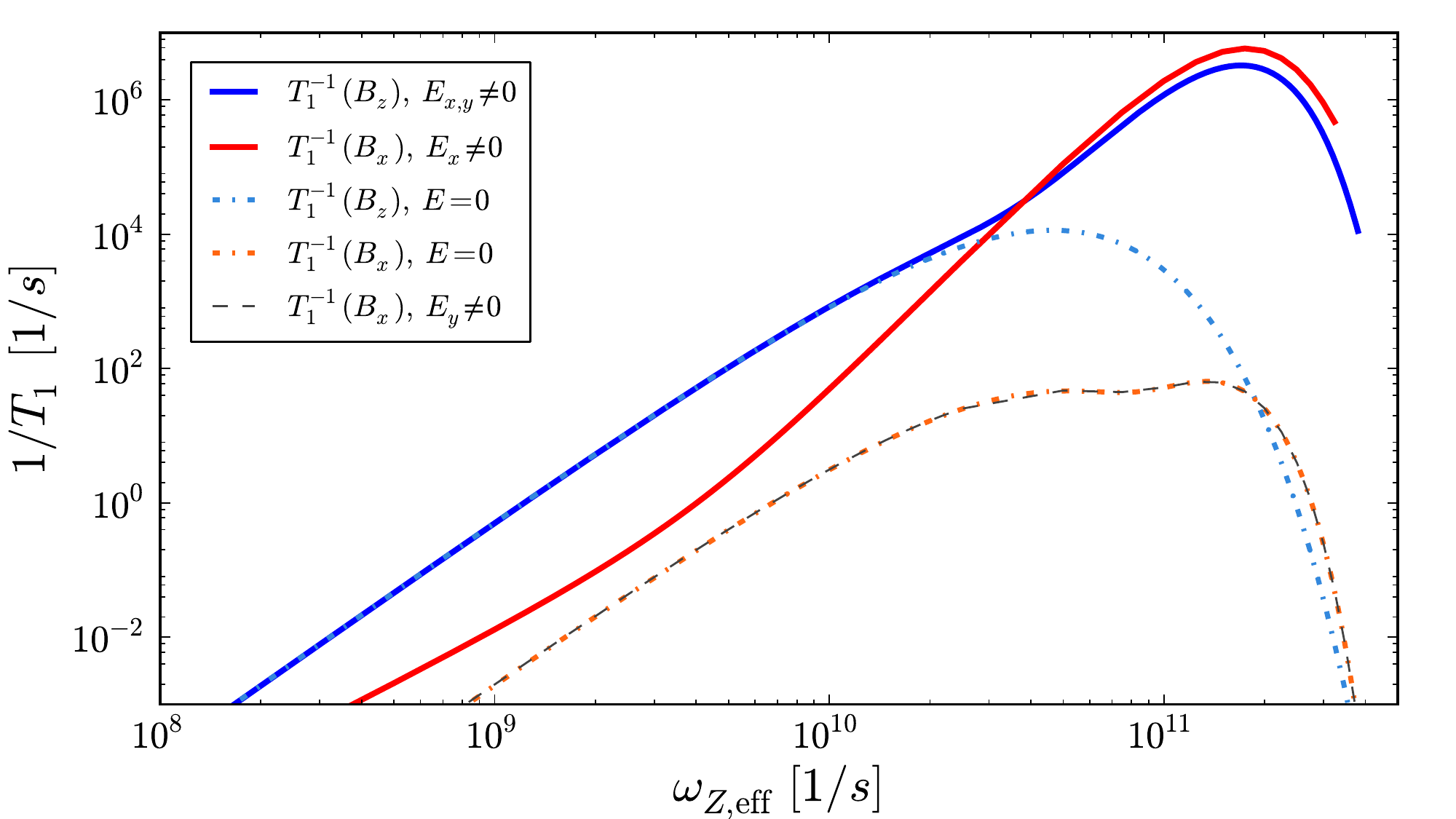} 
\caption{Relaxation rates $T_1^{-1}$ for $\bm{E}\|\hat{\bm{x}}$ with $|\bm{E}| = 0.1 \mbox{ {V}/$\mu${m}}$ for $\bm{B}\|\hat{\bm{x}}$ (red, solid) and $\bm{B}\|\hat{\bm{z}}$ (blue, solid). 
For comparison we plot $T_1^{-1}$ at $|\bm{E}|=0$ (dotted). 
We find maximal values $T_{1,\text{max}}^{-1}(\bm{B}\|\hat{\bm{z}})\approx 3.2\mbox{ $\mu${s}}^{-1}$, $T_{1,\text{max}}^{-1}(\bm{B}\|\hat{\bm{x}})\approx 5.8\mbox{ $\mu${s}}^{-1}$.
Rotating the electric field so that $\bm{E}\|\hat{\bm{y}}$ yields the same curve for $\bm{B}\|\hat{\bm{z}}$.
Remarkably, for $\bm{B}\|\hat{\bm{x}}$ almost no difference between the curves at $\bm{E}\|\hat{\bm{y}}$ (dashed) and $|\bm{E}|=0$ (dotted) is observed. 
We use the NW and QD parameters given below Fig.~\ref{fig:geffective}. 
\label{fig:relaxrate105fields}}
\end{figure}
In Fig.~\ref{fig:relaxratezerofields},  we display $T_1^{-1}$ for $|\bm{E}|=0$ and two different directions of $\bm{B}$ with respect to the wire, $\bm{B}\|\hat{\bm{z}}$ and $\bm{B}\|\hat{\bm{x}}$.
In this case, the spin-phonon coupling $H_{\text{s-ph},\text{eff}}$ depends only on the coupling terms of $H_{B,\text{orb}}$.
For low $\omega_{Z, \text{eff}}$, i.e.\ the long wavelength regime ($q z_g \ll1$), both curves are proportional to $\omega_{Z, \text{eff}}^{7/2}$. 
This behavior is valid for low temperatures $(\hbar \omega_{Z, \text{eff}}\gg k_B T)$ and will be replaced by $T_1^{-1}\propto \omega_{Z, \text{eff}}^{5/2} T$ for $\hbar \omega_{Z, \text{eff}}\ll k_B T$. 
The $\omega_{Z, \text{eff}}^{7/2}$ scaling contrasts the $\omega_{Z, \text{eff}}^5$ behavior of electrons in QDs. \cite{Khaetskii2001,Golovach2004,Trif2008,Kroutvar2004,Amasha2008, Dreiser2008}
For $\bm{B}\|\hat{\bm{z}}$, only the $F_{\pm1}$ modes contribute significantly to $T_1^{-1}$. 
When directing $\bm{B}\|\hat{\bm{x}}$, the $F_{\pm1}$ contributions dominate for low $\omega_{Z, \text{eff}}$ and, for the chosen QD geometry, are replaced by a dominating $T$ contribution at $|\bm{B}|\approx 150\mbox{ mT}$ (Fig.~\ref{fig:relaxratezerofields}, dashed). 
This results in a double peak whose relative height can be modified by changing $z_g$ or $R$ and $R_s$. 
Most remarkably, for $\bm{B}\|\hat{\bm{x}}$, $T_1^{-1}$ is several orders of magnitude smaller than for $\bm{B}\|\hat{\bm{z}}$.
For the chosen QD geometry, $T_1^{-1}$ reaches maximal values $T_{1,\text{max}}^{-1}(\bm{B}\|\hat{\bm{x}})\approx 60\mbox{ s}^{-1}$ and $T_{1,\text{max}}^{-1}(\bm{B}\|\hat{\bm{z}})\approx 11\mbox{ ms}^{-1}$.
These rates are, depending on the direction of $\bm{B}$, comparable to or much smaller than for electrons in InAs NW QDs. \cite{Trif2008}

Considering non-zero electric fields, we plot $T_1^{-1}$ for $\bm{E}\|\hat{\bm{x}}$ again for $\bm{B}\|\hat{\bm{z}}$ and $\bm{B}\|\hat{\bm{x}}$ (Fig.~\ref{fig:relaxrate105fields}). 
We add the corresponding curves for $|\bm{E}|=0$ (Fig.~\ref{fig:relaxrate105fields}, dashed) to allow for comparison. 
For both orientations of $\bm{B}$, $T_1^{-1}$ is enhanced significantly for larger $\omega_{Z,\text{eff}}$.
This is due to phonons of the $L$ mode coupling $|\!\!\Uparrow\rangle$ and $|\!\!\Downarrow\rangle$ via a combination of  $H_{\text{LK}_{\text{od}}}$ and $H_{\text{DR}}$ which dominates $H_{\text{R}}$.
Due to cylindrical symmetry, applying $\bm{E}\|\hat{\bm{y}}$ for $\bm{B}\|\hat{\bm{z}}$ results in the same effect as described for $\bm{E}\|\hat{\bm{x}}$. 
Remarkably, in stark contrast to $\bm{E}\|\bm{B}\|\hat{\bm{x}}$, only minor changes with respect to the curve at $|\bm{E}|=0$ (Fig.~\ref{fig:relaxrate105fields}, dotted) are observed when $\bm{E}\|\hat{\bm{y}}$ and $\bm{B}\|\hat{\bm{x}}$ (Fig.~\ref{fig:relaxrate105fields}, dashed).
In the latter case, the dominant contributions of $H_{\text{s-ph},\text{eff}}$ are already present in $H_{\text{s-ph},\text{eff}}$ for $|\bm{E}|=0$.

In both cases, $|\bm{E}|=0$ and $|\bm{E}|\neq 0$, increasing the relative shell thickness $\gamma$ shifts the $T_1^{-1}$ curves to slightly larger $\omega_{Z,\text{eff}}$ and lowers the peak height, e.g.\ increasing $\gamma$ from $0.3$ to $0.7$ reduces $T_1^{-1}$ by a factor $\simeq 3$. 
However, decreasing (increasing) $R$ and $R_s$ while keeping $\gamma$ constant has no substantial effect on $T_1^{-1}$ aside from slight shifts to the right (left) on the $\omega_{Z,\text{eff}}$ axis.
Additionally, for $|\bm{E}|=0$ and for $\bm{E}\|\hat{\bm{y}}$ and $\bm{B}\|\hat{\bm{x}}$, enhancing the  confinement length $z_g$ lowers $T_{1,\text{max}}^{-1}$ since the short wavelength regime is reached for smaller $\omega_{Z,\text{eff}}$.
This effect is quite large, for instance raising $z_g$ from $60$ to $100\mbox{ nm}$ tunes $T_{1,\text{max}}^{-1}$ by factors between $10$ and $100$.
However, for $\bm{E}\|\hat{\bm{x}}$ and $\bm{B}\|\hat{\bm{z}}$ or $\bm{B}\|\hat{\bm{x}}$, increasing $z_g$ results in larger $T_{1,\text{max}}^{-1}$. 
From this analysis we conclude that there exist optimal configurations of $\bm{B}$ and $\bm{E}$ in order to obtain long $T_1$ in this type of NW QD. 
$\bm{B}$ should be applied perpendicular to the NW and the optional $\bm{E}$ should lie perpendicular to both $\bm{B}$ and the NW. 
For vanishing $\bm{B}$, as pointed out in Ref.~\onlinecite{Hu2012}, two-phonon processes \cite{Trif2009} might become relevant. 

In the Bloch-Redfield framework, the decoherence time is given by $T_2^{-1} = (2 T_1)^{-1}+T_{\varphi}^{-1}$, where $T_{\varphi}$ denotes the dephasing time. \cite{Golovach2004, Borhani2006}
For $|\bm{E}|=0$, we find $T_2=2 T_1$ because the corresponding spectral function is super-ohmic and gives $T_{\varphi}^{-1}=0$.
For $|\bm{E}|\neq0$ the SOI results in a non-zero dephasing term $T_{\varphi}^{-1}\neq0$ and hence $T_2<2 T_1$.

%
%
%
%
%%%%%%%%%%%%%%%%%%%%%%%%%%%%%%%%%%%%%%%%%%%%%%%%%%%%%%%%%%%%%%%
%
%
%
In conclusion, we have examined effective Zeeman splitting and hole spin dynamics for holes in the lowest VB of a Ge/Si core/shell NW QD. 
We reported a highly anisotropic effective $g$ factor which is strongly tunable by applying electric fields. 
We calculated relaxation rates and found configurations of electric and magnetic fields which correspond to very long spin relaxation times.
Furthermore we pointed out that the relative shell thickness and the QD confinement length influence the spin relaxation time. \\

%
%
%%%%%%%%%%%%%%%%%%%%%%%%%%%%%%%%%%%%%%%%%%%%%%%%%%%%%%%%%%%%%%%
%
%
%
We thank Peter Stano for helpful discussions. 
This work has been supported by SNF, NCCR Nano, NCCR QSIT, DARPA, and IARPA.
\appendix
\section{Effective 1D Hamiltonian}
Here we display the effective 1D Hamiltonians used in Eqs.\ (\ref{eq:LeadingOrder}) and (\ref{eq:PerturbationOverview}) of the main text. 
An extended derivation of these terms can be found in Ref.~\onlinecite{Kloeffel2011}. 
We use the basis $\{g_+, g_-, e_+, e_-\}$, where the exact form of the states  $g_\pm(x,y)$ and $e_\pm(x,y)$ is given in Ref.~\onlinecite{Kloeffel2011}.
The diagonal elements of the Luttinger-Kohn (LK) Hamiltonian and the strain induced splitting is combined in
\begin{equation}
H_{\text{LK}_{\text{d}}} + H_\text{strain} = \left(
\begin{array}{cccc}
\frac{\hbar^2 k_z^2}{2 m_g}	&0	&0	&0 \\
0	&\frac{\hbar^2 k_z^2}{2 m_g}	&0	&0\\
0	& 0	&\frac{\hbar^2 k_z^2}{2 m_e}+\Delta	&0	\\
0	&0	&0	&\frac{\hbar^2 k_z^2}{2 m_e}+\Delta
\end{array}
\right),\label{eq:HDot}
\end{equation}
where $\hbar k_z$ is the momentum operator along the wire and $m_g\simeq m_0/(\gamma_1+2\gamma_s)$ and $m_e = m_0/(\gamma_1+\gamma_s)$ are the effective masses along $z$.
Here, $\gamma_1$ and $\gamma_s$ are the Luttinger parameters in spherical approximation and $m_0$ denotes the bare electron mass.
$\Delta \equiv \Delta_{\rm LK} + \Delta_{\rm strain}(\gamma)$ denotes the level splitting between $g_\pm$ and $e_\pm$, $\gamma \equiv (R_s - R)/R$ is the relative shell thickness, and $R$ ($R_s$) is the core (shell) radius.
$\Delta_{\text{LK}}=0.73 \hbar^2/(m_0 R^2)$ and the strain dependent energy splitting can take values $\Delta_{\text{strain}}(\gamma)\simeq0-30\mbox{ meV}$. 
The magnetic field $\bm{B}=(B_x, 0, B_z) \equiv  |\bm{B}|(\sin\theta, 0, \cos\theta)$ interacts via the Zeeman coupling
\begin{equation}
H_{B,Z} = \mu_B
\left(
\begin{array}{cccc}
B_z G		&B_{x} K	&0	&0\\
B_{x}K	&-B_z G	&0	&0\\
0&0& B_z F	&B_{x}M\\
0&0	& B_{x}M&-B_z F
\end{array}
\right),\label{eq:HBZ}
\end{equation}
where we set $B_y = 0$ due to cylindrical symmetry.
Here, $\mu_B$ is the Bohr magneton and the parameters take the numerical values $G=-0.06$, $K=2.89$, $M=2.56$ and $F=1.56$.
From the LK Hamiltonian additional off-diagonal coupling terms arise and we find
\begin{equation}
H_{{\text{LK}}_\text{od}}	 = \left(
\begin{array}{cccc}
0	&0	&0	&-i C k_z \\
0	&0	&-i C k_z 	&0\\
0	& i C k_z	&0	&0	\\
i C k_z	&0	&0	&0
\end{array}
\right),\label{eq:HSOLK}
\end{equation}
with coupling constant $C=7.26 \hbar^2/(m_0 R)$.
To describe interactions with the electric field $\bm{E}=(E_x, E_y,0)=|\bm{E}| (\cos\varphi_{\text{el}},\sin\varphi_{\text{el}},0)$, we introduce the shorthand notation $\tilde{E} = |\bm{E}| e^{i \varphi_{\text{el}}}$. 
The effective conventional Rashba SOI interaction reads
\begin{equation}
H_{\text{R}} = \alpha 
\left(
\begin{array}{cccc}
 0 & -i \tilde{E} k_z T   & \tilde{E} S   & 0 \\
 i \tilde{E}^{*} k_z T  & 0 & 0 & -\tilde{E}^{*} S   \\
\tilde{E}^{*} S   & 0 & 0 & -\frac{3}{4} i \tilde{E}^{*} k_z   \\
 0 & -\tilde{E} S   & \frac{3}{4} i \tilde{E} k_z   & 0 \\
\end{array}
\right)\label{eq:HSOxy}
\end{equation}
with $T=0.98$ and $S=0.36/R$ and $\alpha=-0.4 \mbox{ nm}^2 e$.
Although fully taken into account, $H_{\text{R}}$ turns out to be negligible for the typical parameters and electric fields considered here (see Eq.\ (\ref{eq:compDRSOCRSO})). 
The direct, dipolar coupling to $\bm{E}$ is given by
\begin{equation}
H_{\text{DR}} = e U 
\left(
\begin{array}{cccc}
0	&0	&\tilde{E}	&0\\
0	&0	&0	&-\tilde{E}^{*}\\
\tilde{E}^{*}&0&0&0\\
0	&-\tilde{E}&0&0
\end{array}
\right),\label{eq:HED}
\end{equation}
\\
where the parameter $U=0.15 R$ scales linearly in $R$. 
We note that the parameters $S$ and $U$ of $H_{\text{R}}$ and $H_{\text{DR}}$ are related by
\begin{equation}
\frac{e U}{\alpha S} \simeq -1.1 \frac{R^2}{\mathrm{nm}^2}, \label{eq:compDRSOCRSO}
\end{equation}
hence $H_{\text{DR}}$ dominates $H_{\text{R}}$ by one to two orders of magnitude for radii between $R=5-10\mbox{ nm}$. 
Lastly, the interaction of the magnetic field via the orbital motion is given by
\begin{eqnarray}
H_{B, \text{orb}} &=& i\mu_B
\left(
\begin{array}{cccc}
0&0&-B_{x}L k_z	&- B_z D k_z\\
0&0& B_z D k_z	&-B_{x}L k_z\\
B_{x}L k_z&- B_z D k_z&0&0\\
 B_z D k_z&B_{x}L k_z	& 0&0
\end{array}
\right)\nonumber\\
&&\label{eq:HBorb}
\end{eqnarray}
with $L=8.04 R$ and $D=2.38 R$. 

%
%
%
%%%%%%%%%%%%%%%%%%%%%%%%%%%%%%%%%%%%%%%%%%%%%%%%%%%%%%%%%%%%%%%
%
%
%
\section{Effective 1D phonon Hamiltonian\label{sec:AppEff1Dphonons}}
Starting from the spherical Bir-Pikus Hamiltonian (see Eq.~(\ref{eq:HBPspher}) of the main text), we derive an effective hole phonon coupling Hamiltonian for each phonon mode $\lambda$. 
This is done by integrating out the transverse components of the matrix elements of the states $g_\pm(x,y)$ and $e_\pm(x,y)$.  
To improve readability, we introduce an effective phonon annihilation operator $\tilde{b}_{\lambda} = e^{i q z} b_{q,\lambda}(t) = e^{i q z}e^{-i \omega_{\lambda}(q) t}b_{q,\lambda}$.\\
The transversal phonon mode $T$ couples as 
\begin{eqnarray}
H_{T}&=&\sum_q a_1\left(
\begin{array}{cccc}
0&    0&    0&    \tilde{b}_{T}-\tilde{b}_{T}^{\dagger}\\
0&    0&    -\tilde{b}_{T}+\tilde{b}_{T}^{\dagger}&    0\\
0&   \tilde{b}_{T} -\tilde{b}_{T}^{\dagger}&    0&    0\\
-\tilde{b}_{T}+\tilde{b}_{T}^{\dagger}&    0&    0&    0
\end{array}\right),\nonumber\\
&&\label{eq:HphonT}
\end{eqnarray}
and the dilatational mode $L$ gives
\begin{widetext}
\begin{equation}
H_{L}=\sum_q \left(
\begin{array}{cccc}
a_2(\tilde{b}_{L}+\tilde{b}_{L}^{\dagger})	&0	&0	&i a_3(\tilde{b}_{L}-\tilde{b}_{L}^{\dagger})\\
0	&a_2(\tilde{b}_{L}+\tilde{b}_{L}^{\dagger})	&	i a_3 (\tilde{b}_{L}-\tilde{b}_{L}^{\dagger})	&0\\
0	&i a_3(\tilde{b}_{L}-\tilde{b}_{L}^{\dagger})&		a_4(\tilde{b}_{L} +\tilde{b}_{L}^{\dagger})	&0\\
i a_3(\tilde{b}_{L}-\tilde{b}_{L}^{\dagger})&	0&	0&	a_4(\tilde{b}_{L}+\tilde{b}_{L}^{\dagger})
\end{array}\right).\label{eq:HphonL}
\end{equation}
\end{widetext}
For the two flexural modes $F_{\pm1}$, we find
\begin{equation}
H_{F_{+1}}=\sum_q i a_5\left(
\begin{array}{cccc}
0&0&\tilde{b}_{F_{+1}}^{\dagger}&0\\
0&0&0&\tilde{b}_{F_{+1}}\\
-\tilde{b}_{F_{+1}} &0&0&0\\
0&-\tilde{b}_{F_{+1}}^{\dagger}&0&0
\end{array}\right)\label{eq:Hphon1}
\end{equation}
and
\begin{equation}
H_{F_{-1}}=\sum_q i a_5\left(
\begin{array}{cccc}
0&0&-\tilde{b}_{F_{-1}}&0\\
0&0&0&-\tilde{b}_{F_{-1}}^{\dagger}\\
\tilde{b}_{F_{-1}}^{\dagger}&0&0&0\\
0&\tilde{b}_{F_{-1}}&0&0\end{array}\right).\label{eq:Hphon-1}
\end{equation}
Here, the $a_i$, $i=1,2,3,4,5$, are real, $q$-dependent prefactors. 
The complete effective hole phonon coupling $H_{\text{h-ph}}$ is then given by
\begin{equation}
H_{\text{h-ph}} = \sum_{\lambda}H_{\lambda} = H_{\text{T}}+H_{\text{L}}+H_{F_{+1}}+H_{F_{-1}}.\label{eq:Hhphcomplet}
\end{equation}

\end{document}